\documentclass[]{aa}  
\usepackage{graphicx}
\usepackage{natbib}
\begin{document}

\title{A companion candidate in the gap of the T~Cha transitional disk\thanks{Based on observations 
obtained at the European Southern Observatory using the Very Large Telescope in Cerro Paranal, 
Chile, under program 84.C-0755(A)}}
   \author{
          N. Hu\'elamo \inst{1} 
         \and
          S. Lacour\inst{2}
          \and
          P. Tuthill\inst{3}
          \and
          M. Ireland\inst{3}
          \and
           A. Kraus\inst{4}       
          \and 
          G. Chauvin\inst{5}
   }     
    \offprints{N. Hu\'elamo}
   \institute{
   Centro de Astrobiolog\'{\i}a (INTA-CSIC);  ESAC Campus, P.O. Box 78, E-28691 Villanueva de la Ca\~nada, Spain\\
   \email{nhuelamo@cab.inta-csic.es}
   \and
  Observatoire de Paris, LESIA/CNRS UMR, 5 place Jules Janssen, Meudon, France
   \and 
  Sydney Institute for Astronomy, School of Physics, University of Sydney, NSW 2006, Australia
  \and     
   Hubble Fellow; University of Hawaii-IfA, 2680  Woodlawn Dr, Honolulu, HI 96822, USA
   \and
   Laboratoire d'Astrophysique de Grenoble, CNRS, Universit\'e Joseph Fourier, 
   UMR 5571, BP 53, F-38041 Grenoble, Cedex 9, France
   }
   \date{Received; accepted}

  \abstract
   {T Cha is a young star surrounded by a cold disk.
The presence of a gap within its disk, inferred from fitting to the spectral
energy distribution, has suggested on-going planetary formation.}
   {The aim of this work is to look for very low-mass companions within the disk gap of T Cha.}
   {We observed T Cha in $L'$ and $K_s$ with NAOS-CONICA, the adaptive optics system at the VLT, using sparse aperture masking.}
   {We detected a source in the $L'$ data at a separation of 62$\pm$7\,mas, position angle of $\sim$78$\pm$1 degrees, and
   a contrast of $\Delta L'\, =\, 5.1\pm0.2$\,mag. The object is not detected
   in the $K_s$ band data, which show a 3-$\sigma$ contrast limit of 5.2\,mag at the position of the
   detected $L'$ source. 
   For a distance of 108\,pc, the detected companion candidate is located at 6.7\,AU from the primary, well within the disk gap. If T~Cha and the
   companion candidate are bound,
   the comparison of  the  $L'$ and $K_s$ photometry with  evolutionary tracks shows that the photometry is inconsistent with any
   unextincted photosphere at the age and distance of T Cha. The  detected object shows a very red $K_s-L'$ color for which a possible explanation would be a significant amount of dust around it. This would imply that the companion candidate is young, which would strengthen the case for a physical companion, and moreover that the object would be in the substellar regime, according to the $K_s$ upper limit. 
    Another exciting possibility would be that this companion is a recently formed planet within the disk.  
 Additional observations are mandatory to confirm that the object is bound 
 and to properly characterize it.}
   {}

\authorrunning{Hu\'elamo et al.}
\titlerunning{A companion candidate in the gap of the T Cha disk.} 
\maketitle
%

\section{Introduction}

In recent years, a large number of disks characterized by a lack of
significant mid-infrared (IR) emission and a rise into the far-IR 
have been detected \citep[e.g.][]{Brown2007,Merin2010}. 
These are the so-called `transitional disks', and they are thought to be in an 
intermediate evolutionary state between primordial Class II protoplanetary disks and Class III
debris disks.

The lack of mid-IR excess in cold disks has been interpreted as a sign
of dust clearing, which can result in  gaps or holes
within the disk.  These gaps and holes can be created by several
mechanisms, such as a close stellar companion, disk
photoevaporation, grain growth or a planet formed within the disk. 
A planet forming within the disk is expected to generate a gap while the dust and gas is accreted onto its
surface, sweeping out the orbital region \citep[e.g.][]{Lubow1999}.

In this work, we present high angular resolution deep IR observations 
of T Cha, a young star with a cold disk.  Its spectral energy distribution 
(SED) shows a small IR excess between 1--10\,$\mu$m and a very steep rise
between 10--30\,$\mu$m.  The SED has only been successfully modeled by 
including a gap from 0.2 to 15\, AU \citep{Brown2007,Schisano2009}. In fact, an 
inner dusty disk has recently been detected by \citet[][]{Olofsson2011}. Because 
one of the possibilities is that the gap has been cleared by 
a very low-mass object, we obtained 
adaptive optics (AO) sparse aperture masking (SAM) observations of T~Cha aimed at
detecting faint companions within the disk gap.

\section{The target: T~Cha}

T~Cha is a high probability member of the young $\epsilon$ Cha
association \citep{Torres2008}.  It is a G8-type star with a mass of $\sim$ 1.5\,M$_{\odot}$,
classified as  a weak-lined T Tauri star based on the H$_\alpha$ equivalent width
from single epoch spectroscopy \citep{Alcala1993}.  Subsequent
photometric and spectroscopic monitoring has indicated a strong
variability of this line, which shows significant changes in its
equivalent width, intensity, and profile \citep[][]{GregorioHetem1992,
  Alcala1993, Schisano2009}. If the line is related to accretion
episodes, then the average accretion rate is $\dot{M}$ = 4$\times$
10$^{-9}\,M_{\odot}/yr$ \citep{Schisano2009}.

T~Cha shows variable circumstellar extinction with a most frequent value of $A_V$=1.7\,mag according to \citet{Schisano2009}. 
The authors derive a disk  extinction law characterized by $R_V=5.5$, which suggests 
the presence of large dust  grains within the disk.

The age of the source is variously estimated to be between 2--10\,Myr 
according to different methods \citep{Fernandez2008}. A complete study of the
$\epsilon$ Cha association by \citet{Torres2008} provides an average
age of 6\,Myr, while \citet{daSilva2009} estimate a slightly older age
(between 5-10\,Myr) based on the lithium content of the $\epsilon$ Cha
members.  Finally, a dynamical evolution study of the $\eta$ Cha
cluster, which probably belongs to the $\epsilon$ Cha association,
provides an age of 6.7\,Myr \citep{Ortega2009}.  For the purpose of
this paper, we adopt an age of 7\,Myr.

The distance to the source, based on the Hipparcos parallax, is
66\,pc$\pm$15\,pc.  A more reliable value of 100\,pc was obtained using proper-motion studies
\citep[][]{Frink1998,Terranegra1999}.  \citet{Torres2008} provided a
kinematical distance of 109\,pc for T~Cha, and an average value of
108$\pm$9\,pc for the whole association. We adopted the latter
value for this paper.

Finally, previous works based on radial velocity (RV) and
direct imaging and coronographic techniques have not reported the
presence of any (stellar or very low-mass) companion around T~Cha
\citep[][]{Schisano2009,Chauvin2010,Vicente2011}. The SAM 
observations allow us to fill the gap between  between RV and direct imaging observations.

   \section{Observations and data reduction}

The observations presented here were obtained with NAOS-CONICA (NACO),
the AO system at the Very Large Telescope (VLT), and SAM
\citep{Tuthill2010} in two different campaigns.  The {\em L'} observations
were obtained in March 2010 under excellent atmospheric conditions
(average coherent time of $\tau_0$=8\,ms, and average seeing of
0\farcs6), while the $K_s$ data were obtained in July 2010 under
moderate atmospheric conditions ($\tau_0$=4\,ms, and seeing of
1\farcs0).

In March 2010, T~Cha was observed with the L27 objective, the seven-hole mask and the $L'$
filter ($\lambda_c$= 3.80$\mu$m, $\Delta\lambda$=\,0.62 $\mu$m). The
target and a calibrator star (HD\,102260) were observed during 10
minutes each. We repeated the sequence star+calibrator nine times,
integrating a total of 48 minutes on-source.  The observational
procedure included a dithering pattern that placed the target in the
four quadrants of the detector. We acquired datacubes of 100 frames of
0.4\,sec integration time in each offset position.
The plate scale, 27.10$\pm$0.10\,mas/pix,
and true north orientation of the detector, -0.48$\pm$0.25 degrees, were derived using the astrometric 
calibrator $\theta$ Ori$^1$~C observed in April 2010.

For the $K_s$-band data we used the S27 objective  and the same strategy, but
integrating in datacubes of 100 frames with 0.5\,sec of individual
exposure time. We spent a total of 20 minutes on-source, and we used
two stars, HD\,102260 and HD\,101251, as calibrators.

All data were reduced using a custom pipeline detailed in Lacour et al. 
(in preparation). In brief, each frame is flat-fielded, dark-subtracted, 
and bad-pixel-corrected. The complex amplitudes of each one of the 
$7\times6/2=21$ fringe spatial frequencies were then used to calculate 
the bispectrum, from which the argument is taken to derive the closure phase. 
Lastly, the closure phases were fitted to a model of a binary source.

\begin{figure}[t]
   \centering
   \resizebox{0.9\hsize}{!}{\includegraphics{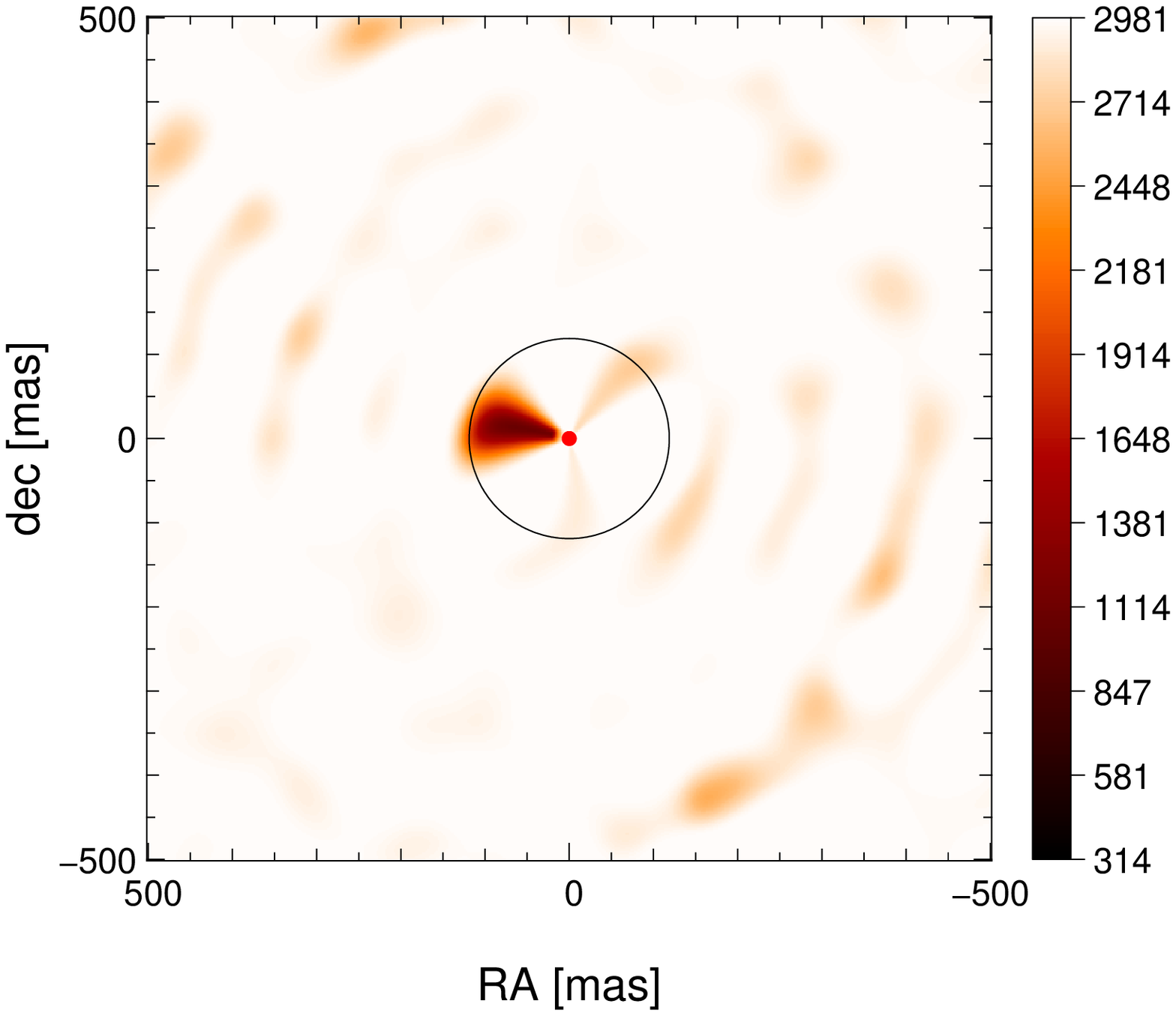}
\includegraphics{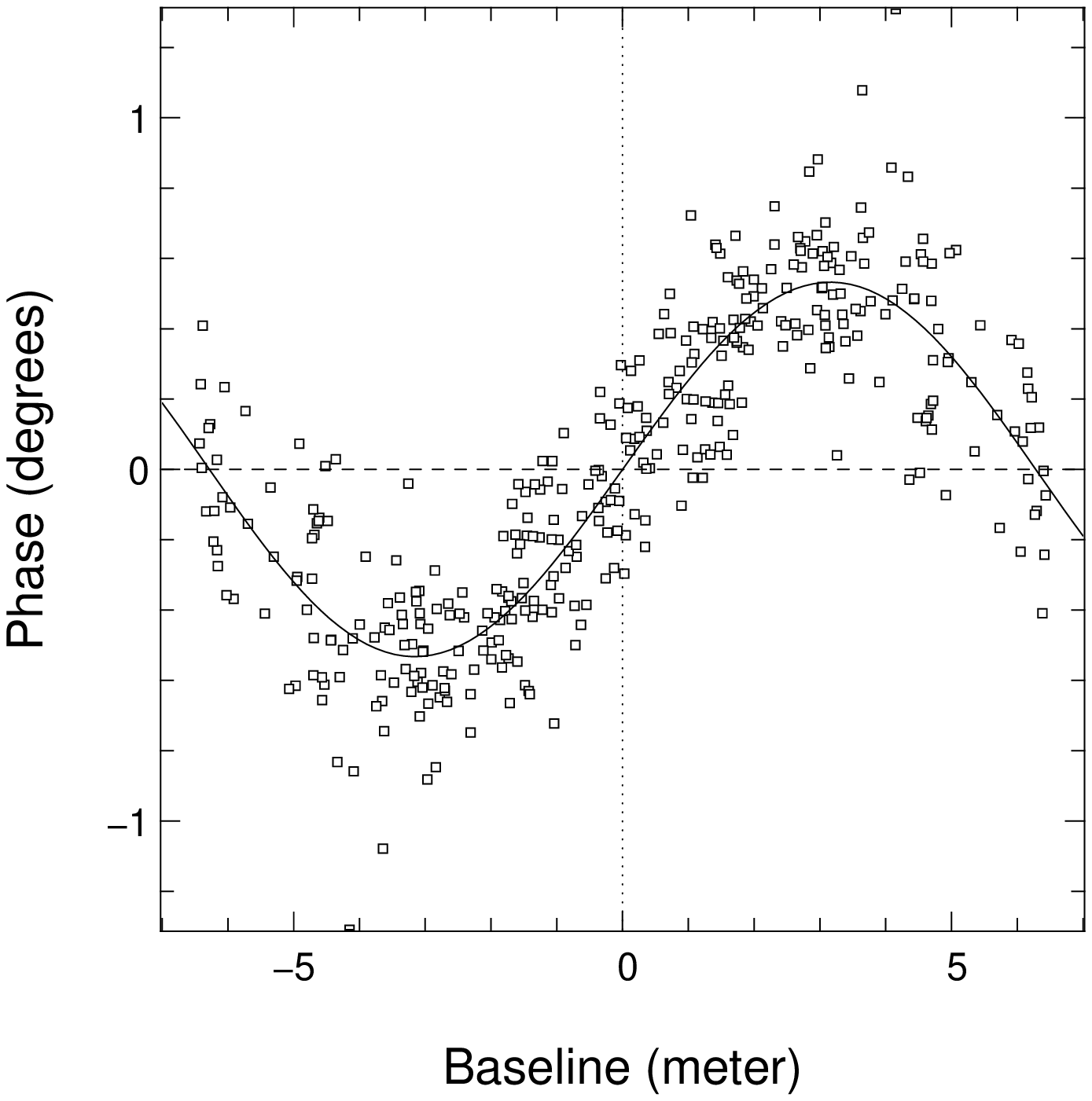}}
\includegraphics[scale=.35]{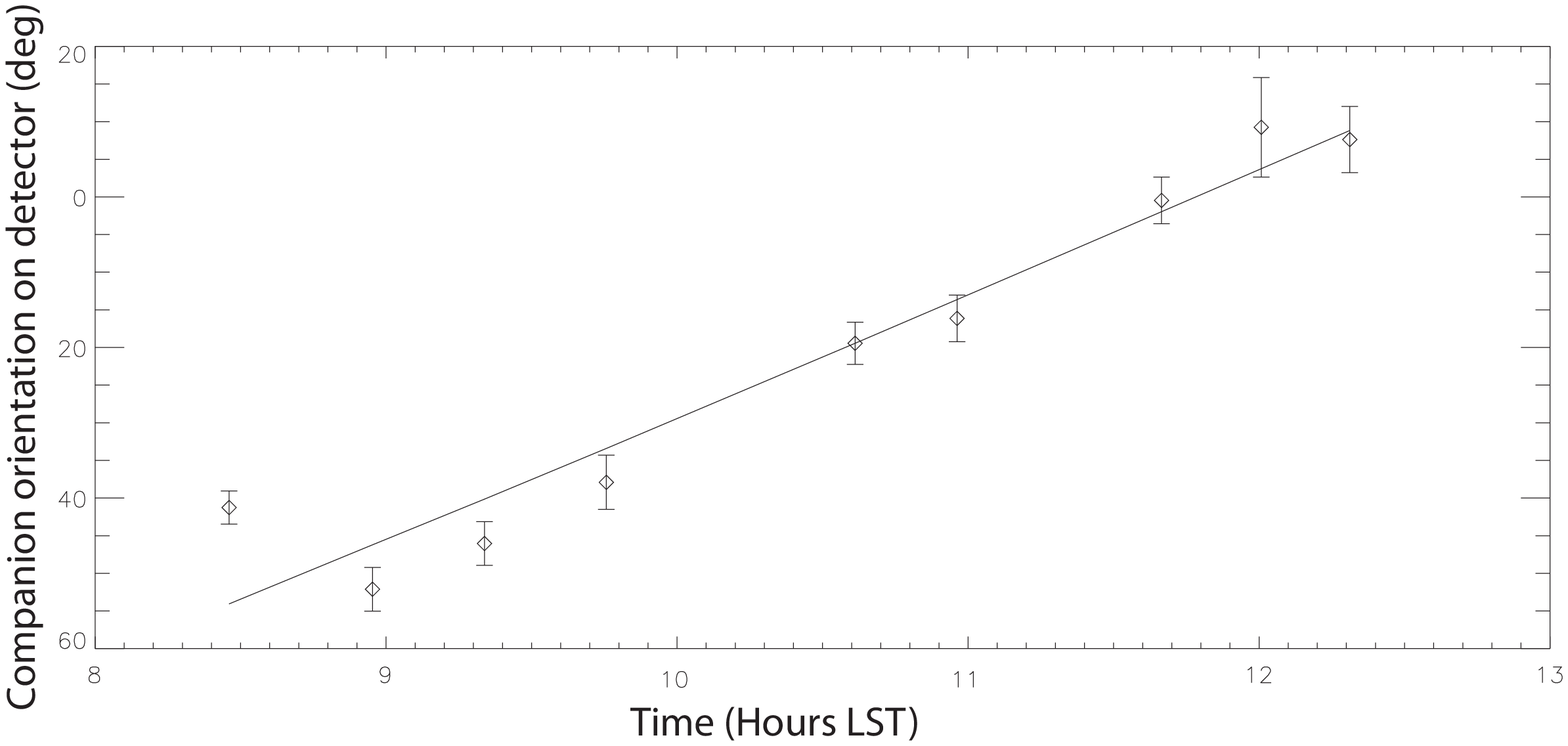}
      \caption{Results from the $L'$ SAM observations: a companion
        candidate is detected at $61.8\pm 7.4$\,mas from the
        central source with a flux ratio of 0.92$\pm$0.20\,\% in
        {\em L'}. {\em Upper left panel:} $\chi^2$ as a function of the
        position of companion candidate (degrees of freedom = 314). The
        black circle corresponds to the imaging resolution of the
        telescope ($1.22\lambda/D$). {\em Upper right panel:} best fit of
        the model of the companion (solid line) overplotted on the
        deconvolved phase. 
        {\em Lower panel:} Orientation of companion detections made using
        each of the nine individual data files plotted in raw detector coordinates. 
        Spurious structures should appear fixed, while real features will 
        rotate with the sky, as illustrated by the overplotted solid line that depicts
        the expected orientation for an object with a sky position angle of 78 degrees.}
        \label{fig1} %
   \end{figure}

\section{Main results}

\subsection{$L'$ detection} \label{detect}

The three free parameters of the fit are the flux ratio, the
separation, and the position angle 
of the companion candidate.  The upper left panel of Fig.~\ref{fig1} 
depicts the minimum $\chi^2$ as a function of position angle and separation. 
For an arbitrary fit,  $\chi^2$ is high (reduced $\chi^2$ of $\approx 9$), but the 
map shows  a clear minimum for a companion to the west of the star. 
The phase corresponding to the best-fit model
is shown in the right panel of the same figure. It consists of a
sinusoidal curve with a specific angular direction, and a period of
half the resolution of the 8.2 meter telescope. In the same figure we
plotted the deconvolved phases from the measured closure phases that were projected
onto the orientation of the best-fit binary.

The best-fitting companion parameters for the L-band data are
a separation of $61.8 \pm 7.4$\,mas, a position angle  of $78.5\pm 1.2$
degrees, and a fractional flux with respect to the central object of 
$0.92 \pm 0.20$\%. To confirm the validity of the
detection, each one of the nine star+calibrator data pairs was also
analyzed separately. Because the observations were taken in `pupil tracking 
mode', all optical and electronic aberrations should remain at frozen
orientation on the detector, while a real structure on the sky will rotate
with the azimuth pointing of the telescope (close to the sidereal
rate). This expected rotation of the detection is illustrated in 
Fig.~\ref{fig1}, which strongly argues against an instrumental artifact.

\begin{figure}[t]
   \centering
 \includegraphics[scale=0.28]{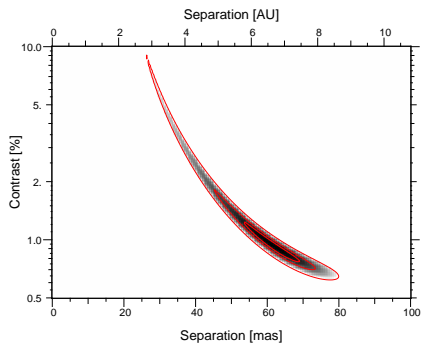}
  \caption{Error contours as a function of separation and flux ratio in the {\em L'} filter. 
  The contour levels correspond to 1-$\sigma$, 2-$\sigma$, and 3-$\sigma$. The upper horizontal 
  labels provide the separation in  astronomical units assuming a distance of 108\,pc.} 
  \label{fig2} 
   \end{figure}

The detection error bars reported above are 1-$\sigma$, but owing
to the low separation, there is a strong degeneracy between separation
and flux ratio. This is highlighted in the contours shown in
Fig.~\ref{fig2}. The limits of the 3~$\sigma$ contours correspond to a
spread of parameters between flux ratio of 9\% at 26\,mas, and 0.6\% at 80\,mas.

\subsection{$K_s$ upper limit}

\begin{figure}[t]
   \centering
   \includegraphics[scale=.28]{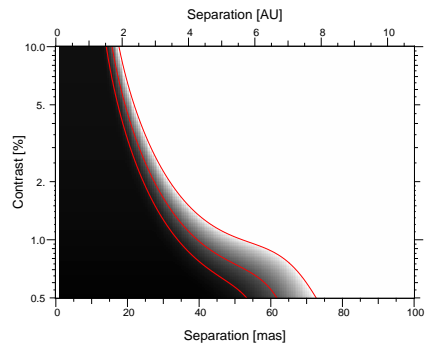}
  \caption{Contrast in the $K_s$-band data. The solid red lines represent the 1-$\sigma$, 
  2-$\sigma$, and 3-$\sigma$ upper limits at different separations from the central source. 
  We estimate a 3-$\sigma$ upper limit of 0.83\% at 62\,mas, that is, at the position of the $L'$ detected source.} 
  \label{fig3} 
   \end{figure}

We did not detect any source around T~Cha in the $K_s$-band data. 
Figure~\ref{fig3} shows the 1-$\sigma$, 2-$\sigma$, and
3-$\sigma$ sensitivity curves as a function of the separation to the
central source.  The data analysis shows that we can rule out 
a companion between 40 and 62\,mas, with contrast ratios
varying between 1.3\% and 0.83\%, respectively, at 99\% confidence.
  
\subsection{Physical parameters}

According to Section~\ref{detect}, we have detected a companion
candidate at a separation and position angle of 61.8$\pm$7.4\,mas and
78.5$\pm$1.2 degrees, respectively, and a flux ratio of 0.0092$\pm$0.0020.
Assuming a distance of 108$\pm$9\,pc, the separation between the star
and the companion candidate is $\sim$6.7$\pm$1.0\,AU.
This lies well within the disk gap of T~Cha, according to the
\citet[][0.2-15\,AU]{Brown2007} and \citet[][0.17-7.5\,AU]{Olofsson2011} disk models.

The flux ratio between the primary and the companion translates
into a difference of magnitude of $\Delta L'\, =\,
5.1\pm0.2$\,mag. Because our observational methods are not optimized for
photometry, we instead rely on $L$-band magnitudes for T~Cha from the 
literature.  The reported Johnson L-band ($\lambda_c$=3.45$\mu$m,
$\Delta\lambda$=0.472\,$\mu$m) brightness of T Cha is
5.86$\pm$0.02\,mag \citep{Alcala1993}, while the IRAC channel-1
(3.55$\mu$m, $\Delta\lambda$=0.75$\mu$m) brightness is
5.74$\pm$0.06\,mag. For the purpose of this paper, we assume an
$L'$ magnitude of 5.8$\pm$0.1\,mag for the primary. Our measured contrast
ratio then implies $L'\sim$10.9$\pm$0.2\,mag for the companion candidate.

We can provide a 3-$\sigma$
limit of $\Delta K_s$$>$5.2\,mag at the separation and position angle
of the $L'$ detection for the $K_s$-band data. Because T Cha itself shows
$K_s$=6.95$\pm$0.02\,mag \citep{2MASS}, the limit to the
companion candidate brightness would be $K_s >$12.15\,mag.

The extinction curve derived for T~Cha using optical data seems significantly flatter 
than that from the interstellar medium \citep{Schisano2009}.
However, because the disk extinction curve is not well known in the infrared regime, we corrected the 
observed magnitudes as  a first approximation using the \citet{Mathis1990} extinction curve with $R_V$=5, and assuming $A_V$=1.7\,mag.
We obtained $A_{K_s}$ $\sim$ 0.2\,mag and $A_{L'}$  $\sim$ 0.1\,mag, which results in magnitudes of 
$L'=10.8\pm0.2$ and $K_s$ $>$11.95\,mag.

\section{Discussion}

We have detected a source  within the disk gap of T~Cha.
We will now discuss the possible nature of this object, keeping in mind that
a detection at a single epoch and waveband provides limited information.

\begin{figure}[t]
   \centering
   \includegraphics[scale=.50]{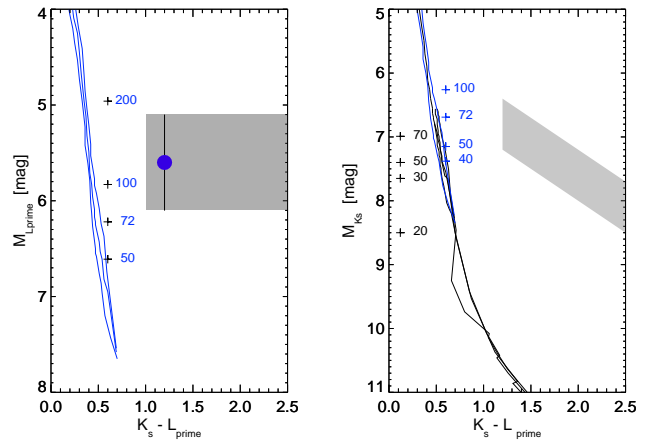}
  \caption{Magnitude-color diagrams with the T Cha companion candidate. {\bf Left:}  The solid lines 
  show NextGEN models for 5,7, and 10\,Myr,  while the shaded area represents the $K_s-L'$ limit. 
  We provide masses (in M$_{Jup}$) for an age of 7\,Myr. {\bf Right:} The solid lines correspond to the 
  NextGEN (blue) and DUSTY (black) models for 5,7, and 10\,Myr, and the masses are derived for an age of 7 Myr. 
  The shaded area shows the possible location of the companion candidate, according to the derived $M_{K_s}$ and $K_s-L'$ limits.} 
  \label{fig4} 
   \end{figure}

If we assume the two objects are co-moving and at a distance of  108$\pm$9\,pc, the 
companion candidate would show absolute magnitudes of $M_{L'}$=\,5.6$\pm$0.5\,mag, and 
$M_{K_s}>$6.8\,mag. We used these values to place the object in two magnitude-color diagrams (Fig. \ref{fig4}).
Assuming they are coeval, and assuming an age for the system of 7$\pm$2\,Myr,
both diagrams show that the observed properties are
inconsistent with any unextincted photosphere at the age and distance of T Cha, according to the NextGen
\citep{Baraffe1998} and DUSTY \citep{Chabrier2000} models.\footnote{We have used
  the latest NextGEN and DUSTY models \citep{allard2010} convolved with the NACO filters.}

The companion candidate shows a very red $K_s-L'$ color. 
A possible explanation for this $K_s-L'$ excess would be a significant amount of dust around the object.
If this scenario is correct, it would imply that the object is young, which would strengthen the case for a physical companion, and moreover
according to Fig.~\ref{fig4}b,  the $K_s$ upper limit would  place it in the substellar (brown dwarf) regime.

Brown dwarf (BD) companions to stars are rare, as we know from radial velocity studies of samples 
of nearby stars \citep[e.g.][]{Grether2006}, giving rise to the so-called
''Brown Dwarf Desert''.
Although this could be interpreted as evidence against a BD for the T~Cha
companion, the separation of 6.7\,AU places it near to the known shores of 
this desert (longer period companions are not well studied).
Indeed, \citet[][]{Kraus2011} claim that intermediate separation
ranges (5--50\,AU) show no evidence for this desert, which makes
T~Cha system interesting whether or not the companion mass lies 
above or below the BD cutoff.

Because disk gaps can be the result of dust clearing owing to planet
formation, we also investigated if the companion candidate could
be a recently formed planet within the disk.
The T Cha system shows
properties that are consistent with this scenario.  First, the
object is detected well within the disk gap.  
The total disk mass derived
by \citet[][]{Olofsson2011}  is 1.74$\pm$0.25$\times$10$^{-2}$\,M$_{\odot}$,
while the average accretion rate is 4$\times10^{-9}$\,$M_{\odot}/yr$.  
These properties seem consistent with a  
planet-forming disk according to \citet{AlexArmi2007}, keeping in mind that both measurements can be affected
by large uncertainties.
If this is the case, the evolutionary models used here are not well suited 
to derive the mass of planetary objects, because we are probably observing 
the planet at the initial formation phase, when the brightness depends only 
on the accretion history and accretion rate.  Indeed, one of the biggest 
advantages of observing transitional disks is that recently formed planets are probably 
still accreting material and therefore should be in their brightest
evolutionary phase.

Additional observations are needed to shed light on the nature of this
exciting object, the first potential substellar object detected within the gap of a transitional disk. 
In particular, observations that detect this object at other wavelengths
or determine the disk position angle and inclination would be most
useful.

\section{Conclusions}

We have observed T Cha with NACO/SAM in two filters, $L'$ and
$K_s$. Our main results can be summarized as follows:

\begin{itemize}

\item We detected a faint companion around T Cha at
  $62\pm7.4$\,mas of separation, position angle of 78.0$\pm$1.2
  degrees, and contrast ratio of $\Delta\,L'=5.04\pm0.2$\,mag.  We
  did not detect any source in the $K_s$-band.  The 3-$\sigma$
  contrast ratio at a separation of 62\,mas is 0.82\%, that is $\Delta
  K_s$$>$5.2\,mag.

\item If T~Cha and the detected object are bound, and assuming a
  distance of 108\,pc, the faint companion lies at 6.7\,AU, that is,
  within the disk gap of the central source.  

\item If T~Cha and its companion are bound and coeval, 
the infrared magnitude-color diagrams show that the object photometry is inconsistent with any unextincted
  photosphere at the age and distance of T~Cha when compared with the evolutionary tracks.

  \item  The companion candidate displays a strong $Ks-L'$ excess, which could be explained by
  a significant amount of dust around it.  This scenario would strengthen the case of a physical companion and, according 
  to the $K_s$ upper limit, it   would place  the object in the substellar regime.

\item The overall properties of the T Cha system also suggest that the newly
  detected source might be a recently formed planet within the disk.
  In this case, suitable planetary formation models are needed to derive its physical properties.

\end{itemize}

Second epoch observations in different photometric bands are mandatory
to confirm if the object is bound, and to properly
characterize it.

\begin{acknowledgements}

This research has been funded by Spanish grants
MEC/ESP2007-65475-C02-02, MEC/Consolider-CSD2006-0070, and
CAM/PRICIT-S2009ESP-1496. We thank the Paranal staff, in particular
A.~Bayo and F.~Selman, for their support during the observations.  NH
is indebted to  D.Stamatellos, H.~Bouy, J. Olofsson, and B.~Mer\'{\i}n for 
useful discussions.
 
\end{acknowledgements}

\bibliographystyle{aa}
\bibliography{16395}

\end{document}